\documentstyle[amstex,amssymb,aps,preprint]{revtex}

\newtheorem{theorem}{Theorem}

\newtheorem{definition}[theorem]{Definition}

\begin{document}
\title{The embedding of the spacetime in five dimensions: an extension of
Campbell-Magaard theorem}
\author{F. Dahia and C. Romero}
\address{Departamento de F\'{\i}sica, Universidade Federal da Para\'{\i}ba\\
C. Postal 5008, Jo\~{a}o Pessoa, PB\\
58059-970, Brazil}
\maketitle
\pacs{04.50+h, 04.20Jb, 98.80Cq}

\begin{abstract}
We extend Campbell-Magaard embedding theorem by proving that any
n-dimensional semi-Riemannian manifold can be locally embedded in an
(n+1)-dimensional Einstein space. We work out some examples of application
of the theorem and discuss its relevance in the context of modern
higher-dimensional spacetime theories.
\end{abstract}

\section{Introduction}

The old idea that our universe is fundamentally higher-dimensional with $%
n=4+d$ spacetime dimensions seems to be gaining grounds very rapidly in
recent years. Mathematical schemes by which our ordinary spacetime is viewed
as a hypersurface embedded in a higher dimensional space have been
considered in several different contexts, such as strings \cite{string},
D-branes \cite{brane}, Randall-Sundrum \cite{randall} models and
non-compactified versions of Kaluza-Klein theories\cite{wesson}.

On the other hand, local isometric embeddings of Riemannian manifolds have
long been studied in differential geometry. Of particular interest is a well
known theorem which states that if the embedding space is flat, then the
minimum number of extra dimensions needed to analytically embed a
n-dimensional Riemannian manifold is $d,$ with $0\leq d\leq n(n-1)/2$ \cite
{einsenhart}.

It turns out, however, that if the embedding space is allowed to be
Ricci-flat then the minimum number of extra dimensions that are necessary
for the embedding falls dramatically to $d=1.$ This is the content of a
little known but powerful theorem due to Campbell \cite{campbell}, the proof
of which was given by Magaard\cite{magaard}. Campbell-Magaard 's result has
acquired fundamental relevance for granting the mathematical consistency of
five-dimensional embedding theories and also has been applied to investigate
how lower-dimensional theories could be related to (3+1)-dimensional vacuum
Einstein gravity\cite{romero}.

The increasing attention given to the Randall-Sundrum model\cite{randall} in
which the embedding space, i.e., the bulk, corresponds to an Einstein space,
rather than a Ricci-flat one, has led us to wonder whether Campbell-Magaard
theorem could be generalized and what sort of generalization could be done.
Research in this direction, where a scheme for extending Campbell-Magaard
theorem for embedding spaces with a non-null Ricci tensor has been put
forward quite recently by Anderson and Lidsey \cite{lidsey}.

The purpose of the present paper is to prove that Campbell-Magaard theorem
can, indeed, be extended to include Einstein spaces. Our proof is entirely
inspired in Magaard 's reasoning although some adaptations to the more
general semi-Riemannian character of the spaces had to be made.

The paper is organized as follows. Section $2$ is devoted to state and prove
an extension of Campbell-Magaard theorem, in which the embedding manifold is
an Einstein space. The proof is rather involved and resorts to auxiliary
theorems and lemmas. In section 3 we apply the general result to some
examples and, finally, Section 4 contains our conclusion.

\section{Isometric embedding in an Einstein space}

Campbell-Magaard theorem for local isometric embedding in Ricci-flat space
refers to Riemannian manifolds, i.e., those endowed with positive-definite
metrics. It turns out that for our purposes of generalization this
restrictive condition is not essential, so in what follows we shall consider
semi-Riemannian manifold with metrics of indefinite signature instead. First
let us introduce some definitions, set the notation and present preliminary
theorems and lemmas.

A n-dimensional manifold $M^{n}$ is termed semi-Riemannian if it is endowed
with a metric, i.e., a symmetric and non-degenerated second-rank tensor
field of arbitrary signature. (In this paper we are considering only
manifolds and metrics which are analytic).

\begin{definition}
Consider a differential map $\Phi :U\subset M^{n}\rightarrow N^{n+k},$ where 
$U$ is an open subset of $M^{n},$ and $N^{n+k}$ $\left( k\geq 0\right) $ is
a manifold of dimension $n+k.$ Then $\Phi $ is called a local isometric
embedding if the following conditions hold:

i) $d\Phi _{p}:T_{p}M^{n}\rightarrow T_{\Phi \left( p\right) }N^{n+k}$ is
injective for all $p\in U;$

ii) $g_{p}(v,w)=\tilde{g}_{\Phi \left( p\right) }\left( d\Phi \left(
v\right) ,d\Phi \left( w\right) \right) $ for all $v,w\in T_{p}M^{n},$ where 
$\tilde{g}$ denotes the metric of $N^{n+k}.$

iii) $\Phi $ is a homeomorphism onto its image in the induced topology.
\end{definition}

If $\Phi $ is of class $C^{k}\left( analytic\right) $ then the embedding is
said to be of class $C^{k}\left( analytic\right) .$ Naturally, a local
isometric embedding may be characterized in terms of coordinates. For
instance, let ${\bf x=}\left\{ x^{1},...,x^{n}\right\} $ and ${\bf y=}%
\left\{ y^{1},...,y^{n+k}\right\} $ denote coordinate patches for $U\subset
M^{n}$ and $V\subset N^{n+k},$ respectively, with $\Phi \left( p\right) \in
V.$ The embedding $\Phi $ determines a relation between the coordinates
denoted by 
\begin{equation}
y^{\alpha }=\sigma ^{\alpha }\left( x^{1},...,x^{n}\right) ,
\end{equation}
where $\sigma ={\bf y}\circ \Phi \circ {\bf x}^{-1}:{\bf x}\left( U\right)
\subset {\Bbb R}^{n}\rightarrow {\Bbb R}^{n+k}.$ (Throughout Latin and Greek
indices will run from $1$ to $n$ and $1$ to $n+k,$ respectively)

In this way the isometric condition leads to 
\begin{equation}
g_{ij}=\frac{\partial \sigma ^{\alpha }}{\partial x^{i}}\frac{\partial
\sigma ^{\beta }}{\partial x^{j}}\tilde{g}_{\alpha \beta }  \label{iso}
\end{equation}
where the functions $g_{ij}$ and $\tilde{g}_{\alpha \beta }$ are the
components of metric with respect to the coordinate bases $\left\{ \partial
_{x^{i}}\right\} $ and $\left\{ \partial _{y^{\alpha }}\right\} ,$
respectively. We, therefore, say that $M^{n}$ can be local and isometrically
embedded in $N^{n+k}$ if there exist $n+k$ differentiable functions $%
y^{\alpha }=\sigma ^{\alpha }\left( x^{1},...,x^{n}\right) ,$ (embedding
functions) such that the Jacobian matrix $\frac{\partial \sigma ^{\alpha }}{%
\partial x^{i}}$ has rank $n$ and (\ref{iso}) holds.

Given two arbitrary semi-Riemannian manifolds $M^{n}$ and $N^{n+k}$ it may
happen that there exists no isometric embedding between them. Thus, it is of
interest to find out conditions assuring or not the existence of embedding,
in particular, if the embedding space $N^{n+k}$ is not specified except that
its belongs to a collection of manifolds, ${\cal M}_{\pi },$whose members,
say, share a certain geometrical property. For example, $\pi $ may express a
restriction of the following kinds: to be flat, to have constant curvature,
to be Ricci-flat, to be an Einstein space, and so forth. This way of
formulating the problem motivates the definition below.

\begin{definition}
We say that a semi-Riemannian manifold $M^{n}$ has an embedding in the set $%
{\cal M}_{\pi },$ if there is at least a member of ${\cal M}_{\pi },$ say $%
N^{n+k},$ in which $M^{n}$ is embeddable.
\end{definition}

The following is a theorem which establishes necessary and sufficient
conditions for the existence of local isometric embedding of a n-dimensional
semi-Riemannian space $\left( M^{n},g\right) $ in the set ${\cal M}_{\pi
}^{n+1}$ of $\left( n+1\right) $-dimensional semi-Riemannian spaces $\left(
N^{n+1},\tilde{g}\right) $ that satisfy the (nonspecified) property $\pi .$
The original version of this theorem is due to Magaard \cite{magaard} and is
restricted to the Riemannian case, though the extension to semi-Riemannian
manifold is straightforward.

{\bf Theorem 1. }{\it Let} $\left( M^{n},g\right) $ {\it constitute a
semi-Riemannian manifold, }${\cal M}_{\pi }^{n+1}=\{\left( N^{n+1},\tilde{g}%
\right) ,$ {\it which satisfy the property} $\pi \}${\it \ and }${\bf x}%
=\left\{ x^{1},...,x^{n}\right\} ${\it \ a coordinate system covering a
neighborhood }$U${\it \ of }$p\in M^{n}.${\it \ A necessary and sufficient
condition for a local analytical embedding of }$M^{n}$ {\it at} $p,$ {\it \
with line element } 
\begin{equation}
ds^{2}=g_{ik}dx^{i}dx^{k},  \label{ds2n}
\end{equation}
{\it in }${\cal M}_{\pi }^{n+1}${\it \ is that}

{\it i) there exist analytic functions} 
\begin{eqnarray}
\bar{g}_{ik} &=&\bar{g}_{ik}\left( x^{1},...,x^{n},x^{n+1}\right) 
\label{gbar} \\
\bar{\phi} &=&\bar{\phi}\left( x^{1},...,x^{n},x^{n+1}\right) 
\label{psibar}
\end{eqnarray}
{\it defined in some open set }$D\subset {\bf x}\left( U\right) \times {\Bbb %
R}${\it \ containing the point }$\left( x_{p}^{1},...,x_{p}^{n},0\right) $ 
{\it and satisfying the conditions} 
\begin{equation}
\bar{g}_{ik}\left( x^{1},...,x^{n},0\right) =g_{ik}\left(
x^{1},...,x^{n}\right)   \label{gbar=g}
\end{equation}
{\it in an open set }$A\subset {\bf x}\left( U\right) ;$%
\begin{eqnarray}
\bar{g}_{ik} &=&\bar{g}_{ki}  \label{ik=ki} \\
\left| \bar{g}_{ik}\right|  &\neq &0  \label{det} \\
\bar{\phi} &\neq &0  \label{psibar/=0}
\end{eqnarray}
{\it in }$D.${\it \ ( }$\left| \bar{g}_{ik}\right| ${\it \ denotes the
determinant of }$\bar{g}_{ik}${\it \ );}

{\it \ ii) and also that} 
\begin{equation}
ds^{2}=\bar{g}_{ik}dx^{i}dx^{k}+\varepsilon \bar{\phi}^{2}dx^{n+1}dx^{n+1},
\end{equation}
{\it where }$\varepsilon ^{2}=+1,${\it \ be a line element in a certain
coordinate neighborhood }$V$ {\it \ of some manifold }$N^{n+1}\in {\cal M}%
_{\pi }^{n+1}.$

The essential idea of this theorem is that there exists a coordinate system
``adapted'' to the embedding in such a manner that the image of the
embedding coincide with the hypersurface $x^{n+1}=0$ of the embedding space
and the condition of isometry reduces to (5).

While the sufficient condition is easily demonstrated, the proof of the
necessary condition of this theorem is very long and will be omitted \cite
{magaard,fdahia}.

We now consider a $\left( n+1\right) -$dimensional semi-Riemannian manifold $%
\left( N^{n+1},\tilde{g}\right) $ and $\left\{ y^{1},...,y^{n+1}\right\} $ a
coordinate system defined in an open set of $V\subset N^{n+1}.$ Let $\tilde{g%
}_{\alpha \beta }$ and $\tilde{R}_{\alpha \beta }$ denote the components of
the metric and Ricci tensor, respectively. The manifold $\left( N^{n+1},%
\tilde{g}\right) $ is called an Einstein space if 
\begin{equation}
\tilde{R}_{\alpha \beta }=\frac{2\Lambda }{1-n}\tilde{g}_{\alpha \beta },
\label{einstein}
\end{equation}
where $n\geq 2$ and $\Lambda $ is a constant. Of course (\ref{einstein}) is
equivalent to $\tilde{G}_{\alpha \beta }=\Lambda \tilde{g}_{\alpha \beta },$ 
$\tilde{G}_{\alpha \beta }$ being the components of the Einstein tensor, and
for this reason $\Lambda $ will be occasionally referred to as the
cosmological constant.

As is well known, at each point of an arbitrary semi-Riemannian space $%
N^{n+1}$ there exists a coordinate neighborhood in which the metric has the
form 
\begin{equation}
ds^{2}=\bar{g}_{ik}dy^{i}dy^{k}+\varepsilon \bar{\phi}^{2}\left(
dy^{n+1}\right) ^{2}.  \label{ds2n+1}
\end{equation}

Let us now consider the inclusion map $\iota \left( y^{1},...,y^{n}\right)
=\left( y^{1},...,y^{n},0\right) .$ This map determines an embedding of the
hypersurface $\Sigma _{0},$ defined by $y^{n+1}=0,$ in $N^{n+1}.$ This
hypersurface endowed with the induced metric by the inclusion map, which is
given by 
\begin{equation}
g_{ik}\left( y^{1},...,y^{n}\right) =\frac{\partial \iota ^{\alpha }}{%
\partial y^{i}}\frac{\partial \iota ^{\beta }}{\partial y^{k}}\tilde{g}%
_{\alpha \beta }=\bar{g}_{ik}\left( y^{1},...,y^{n},0\right) 
\end{equation}
constitutes a semi-Riemannian space. The intrinsic curvature of $\Sigma _{0}$
and the curvature of $N^{n+1}$ calculated at $\Sigma _{0}$ are related by
the Gauss-Codazzi equations. In the coordinates (\ref{ds2n+1}), the
Gauss-Codazzi relations can be written in the form\cite{einsenhart}: 
\begin{eqnarray}
R_{mkij} &=&\tilde{R}_{mkij}+\varepsilon \left( \Omega _{ik}\Omega
_{jm}-\Omega _{jk}\Omega _{im}\right) , \\
\nabla _{j}\Omega _{ik}-\nabla _{i}\Omega _{jk} &=&\frac{1}{\phi }\tilde{R}%
_{\left( n+1\right) kij},
\end{eqnarray}
where $R_{mkij}$ and $\tilde{R}_{mkij}$ \footnote{$R_{kij}^{l}=\Gamma
_{ik,j}^{l}-\Gamma _{jk,i}^{l}+\Gamma _{jm}^{l}\Gamma _{ik}^{m}-\Gamma
_{im}^{l}\Gamma _{jk}^{m}$} are the components of the curvature tensor of $%
\Sigma _{0}$ and of $N^{n+1},$ respectively, $\nabla _{i}$ denoting the
covariant derivative with respect to the metric $g_{ik},$ and $\Omega _{ik}$
(the covariant components of the extrinsic curvature tensor of $\Sigma _{0})$
are given by 
\begin{equation}
\Omega _{ik}=-\frac{1}{2\bar{\phi}}\frac{\partial \bar{g}_{ik}}{\partial
y^{n+1}},  \label{hik}
\end{equation}
in the coordinates (\ref{ds2n+1}). We are now interested in obtaining the
components of the Ricci tensor $R_{\alpha \beta }=g^{\delta \gamma
}R_{\delta \alpha \gamma \beta }$ in the coordinates (\ref{ds2n+1}). With
the help of the Gauss-Codazzi equations and from the expression 
\[
\tilde{R}_{\quad i\left( n+1\right) k}^{\left( n+1\right) }=\left( -\frac{%
\varepsilon }{\phi }\frac{\partial \Omega _{ik}}{\partial y^{n+1}}+\frac{1}{%
\phi }\nabla _{i}\nabla _{k}\phi -\varepsilon g^{jm}\Omega _{jk}\Omega
_{im}\right) ,
\]
which can be obtained by a straightforward calculation, we are left with 
\begin{eqnarray}
\tilde{R}_{ik} &=&R_{ik}+\varepsilon g^{jm}\left( \Omega _{ik}\Omega
_{jm}-2\Omega _{jk}\Omega _{im}\right) -\frac{\varepsilon }{\phi }\frac{%
\partial \Omega _{ik}}{\partial y^{n+1}}+\frac{1}{\phi }\nabla _{i}\nabla
_{k}\phi  \\
\tilde{R}_{i\,\left( n+1\right) } &=&\phi g^{jk}\left( \nabla _{j}\Omega
_{ik}-\nabla _{i}\Omega _{jk}\right)  \\
\tilde{R}_{\left( n+1\right) \,\left( n+1\right) } &=&\varepsilon \phi
^{2}g^{ik}\left( -\frac{\varepsilon }{\phi }\frac{\partial \Omega _{ik}}{%
\partial y^{n+1}}+\frac{1}{\phi }\nabla _{i}\nabla _{k}\phi -\varepsilon
g^{jm}\Omega _{jk}\Omega _{im}\right) .
\end{eqnarray}
At this stage, it should be clear that the results just obtained may be
carried over to any hypersurface $\Sigma _{c}$ defined by $y^{n+1}=c=const.$
Owing to this we shall henceforth introduce a slight different notation: the
metric induced on any $\Sigma _{c},$ i.e., $\bar{g}_{ik}\left(
y^{1},...,y^{n},c\right) ,$ will be denoted simply by $\bar{g}_{ik};$
likewise all quantities associated with $\bar{g}_{ik}$ will be marked with a
bar sign. However, for convenience, only the metric induced on $\Sigma _{0}$
may also be denoted by $g_{ik}$ (without the bar).

Now let us turn our attention to the case when the embedding manifold $%
N^{n+1}$ is an Einstein space. Since we can assume that any point $q\in
N^{n+1}$ lies in some hypersurface $\Sigma _{c},$ we can use the equations
above for decomposing the Ricci tensor of the embedding manifold in terms of
its intrinsic and extrinsic parts with respect to any hypersurface $\Sigma
_{c},$ with $c=y_{q}^{n+1}.$ If $N^{n+1}$ is an Einstein space, then it
follows from (10) that 
\begin{eqnarray}
\tilde{R}_{ik} &=&\bar{R}_{ik}+\varepsilon \bar{g}^{jm}\left( \bar{\Omega}%
_{ik}\bar{\Omega}_{jm}-2\bar{\Omega}_{jk}\bar{\Omega}_{im}\right) -\frac{%
\varepsilon }{\bar{\phi}}\frac{\partial \bar{\Omega}_{ik}}{\partial y^{n+1}}+%
\frac{1}{\bar{\phi}}\bar{\nabla}_{i}\bar{\nabla}_{k}\bar{\phi}=\frac{%
2\Lambda }{1-n}\bar{g}_{ik}  \label{2Rik} \\
\tilde{R}_{i\,\left( n+1\right) } &=&\bar{\phi}\bar{g}^{jk}\left( \bar{\nabla%
}_{j}\bar{\Omega}_{ik}-\bar{\nabla}_{i}\bar{\Omega}_{jk}\right) =0
\label{2Rn+1} \\
\tilde{G}_{\quad \left( n+1\right) }^{\left( n+1\right) } &=&-\frac{1}{2}%
\bar{g}^{ik}\bar{g}^{jm}\left( \bar{R}_{ijkm}+\varepsilon \left( \bar{\Omega}%
_{ik}\bar{\Omega}_{jm}-\bar{\Omega}_{jk}\bar{\Omega}_{im}\right) \right)
=\Lambda   \label{Gn+1}
\end{eqnarray}
where in the last equation, from the definition of the Einstein tensor, $%
\tilde{G}_{\quad \left( n+1\right) }^{\left( n+1\right) }=\tilde{R}_{\quad
\left( n+1\right) }^{\left( n+1\right) }-\frac{1}{2}\tilde{R},$ $\tilde{R}$
being the curvature scalar $\tilde{R}=\tilde{g}^{\alpha \beta }\tilde{R}%
_{\alpha \beta }.$

After writing the equations above let us concentrate on the problem of
embedding a semi-Riemannian manifold in an Einstein space. The equations (%
\ref{2Rik}), (\ref{2Rn+1}) and (\ref{Gn+1}) may be looked upon as a set of
partial differential equations for $\bar{g}_{ik}$ and $\bar{\phi}.$ At this
point, our strategy is to show that if the components $g_{ik}\left(
x^{1},..,x^{n}\right) $ of a metric of $M^{n}$ with respect to some
coordinate system are given, then there exists an open set of ${\Bbb R}^{n+1}
$ where the above-mentioned equations admit a solution $\bar{g}_{ik}\left(
y^{1},...,y^{n},y^{n+1}\right) $ and $\bar{\phi}\left(
y^{1},..,y^{n},y^{n+1}\right) $ which satisfies the initial condition $\bar{g%
}_{ik}=g_{ik}$ at $\Sigma _{0}.$ Moreover, it will be shown that the
functions $\bar{g}_{ik}$ and $\bar{\phi}$ possess all properties which are
necessary to constitute a line element of a (n+1)-dimensional
semi-Riemannian manifold. Then, as $\bar{g}_{ik}$ and $\bar{\phi}$ satisfy
the Eqs. (\ref{2Rik}), (\ref{2Rn+1}) and (\ref{Gn+1}) the metric originated
by them represents that of an Einstein space. Then, by virtue of Theorem 1,
the existence of the embedding will be guaranteed. However, before
proceeding to the final demonstration we shall make use of two lemmas. Let
us consider the first one.

{\bf Lemma 1}. {\it Let the functions }$\bar{g}_{ik}$ {\it and }$\bar{\phi}$%
{\it \ be analytic at }$\left( 0,...,0\right) \in \Sigma _{0}\subset {\Bbb R}%
^{n+1}$ {\it and satisfy the conditions (\ref{ik=ki}), (\ref{det}), (\ref
{psibar/=0}), and the equation (\ref{2Rik}) in an open set of }${\Bbb R}%
^{n+1}${\it \ which contains }$\left( 0,...,0,0\right) \in {\Bbb R}^{n+1}.$ 
{\it If, in addition, }$\bar{g}_{ik}$ {\it and }$\bar{\phi}$ {\it satisfy (%
\ref{2Rn+1}) and (\ref{Gn+1}) at }$\Sigma _{0},${\it \ then }$\bar{g}_{ik}$ 
{\it e }$\bar{\phi}$ {\it also satisfy (\ref{2Rn+1}) and (\ref{Gn+1}) in
some open set of }${\Bbb R}^{n+1}${\it \ containing }$\left(
0,...,0,0\right) .$

{\it Proof.} By assumption $\bar{g}_{ik}$ and $\bar{\phi}$ satisfy (\ref
{ik=ki}), (\ref{det}) and (\ref{psibar/=0}), hence the functions $\tilde{g}%
_{\alpha \beta }$ defined by $\tilde{g}_{ik}=\bar{g}_{ik}$, $\tilde{g}%
_{n+1n+1}=\bar{\phi}^{2}$ and $\tilde{g}_{in+1}=0$ for $i,k=1,...n,$ may be
considered as the components of a (n+1)-dimensional metric tensor $\tilde{g}$%
. The coefficients of the connection and the components of the curvature
tensor associated to the metric $\tilde{g}_{\alpha \beta }$ can be
calculated as usual. Let us now define the tensor $\tilde{F}_{\alpha \beta }=%
\tilde{G}_{\alpha \beta }-\Lambda \tilde{g}_{\alpha \beta },$ where $\tilde{G%
}_{\alpha \beta }$ is the Einstein tensor. Then, as a consequence of Bianchi
identities for the curvature tensor, it follows that $\tilde{F}_{\alpha
\beta }$ is divergenceless, i.e., 
\begin{equation}
\tilde{\nabla}_{\alpha }\tilde{F}_{\beta }^{\alpha }=0.
\end{equation}
This equation can be rewritten as 
\begin{equation}
\frac{\partial \tilde{F}_{\beta }^{n+1}}{\partial y^{n+1}}=-\frac{\partial 
\tilde{F}_{\beta }^{i}}{\partial y^{i}}-\tilde{\Gamma}_{\mu \lambda }^{\mu }%
\tilde{F}_{\beta }^{\lambda }+\tilde{\Gamma}_{\lambda \beta }^{\mu }\tilde{F}%
_{\mu }^{\lambda }.  \label{divF}
\end{equation}
We have assumed that $\bar{g}_{ik}$ and $\bar{\phi}$ satisfy the equations (%
\ref{2Rik}), (\ref{2Rn+1}) and (\ref{Gn+1}) at $\Sigma _{0},$ hence $\tilde{F%
}_{\beta }^{\alpha }=0$ at $\Sigma _{0}.$ Moreover, $\left. \frac{\partial 
\tilde{F}_{\beta }^{i}}{\partial y^{i}}\right| _{y^{n+1}=0}=\frac{\partial }{%
\partial y^{i}}\left( \left. \tilde{F}_{\beta }^{i}\right|
_{y^{n+1}=0}\right) =0.$ Therefore, we conclude from (\ref{divF}) that 
\begin{equation}
\left. \frac{\partial \tilde{F}_{\beta }^{n+1}}{\partial y^{n+1}}\right|
_{y^{n+1}=0}=0.
\end{equation}

Let us look into Eq. (\ref{divF}){\it \ }separately for $\beta =n+1$ and $%
\beta =i$. Taking first $\beta =n+1$ gives 
\begin{equation}
\frac{\partial \tilde{F}_{n+1}^{n+1}}{\partial y^{n+1}}=-\frac{\partial 
\tilde{F}_{n+1}^{i}}{\partial y^{i}}-\tilde{\Gamma}_{\mu \lambda }^{\mu }%
\tilde{F}_{n+1}^{\lambda }+\tilde{\Gamma}_{\lambda \,n+1}^{n+1}\tilde{F}%
_{n+1}^{\lambda }+\tilde{\Gamma}_{n+1\,n+1}^{i}\tilde{F}_{i}^{n+1}+\tilde{%
\Gamma}_{k\,n+1}^{i}\tilde{F}_{i}^{k}
\end{equation}
From the definition of the Einstein tensor we can write $\tilde{G}_{j}^{i}=%
\tilde{R}_{j}^{i}-\delta _{j}^{i}\left( \tilde{R}_{k}^{k}+\tilde{G}%
_{\,n+1}^{n+1}\right) .$ By assumption, $\tilde{R}_{ij}=\frac{2\Lambda }{1-n}%
\bar{g}_{ij}$ not only at the hypersurface $y^{n+1}=0,$ but also for some
open set $V\subset $ ${\Bbb R}^{n+1}$, with $0\in U.$ Thus, the equality $%
\tilde{F}_{i}^{k}=-\delta _{i}^{k}\tilde{F}_{\,n+1}^{n+1}$ holds in $V.$
This implies that 
\begin{equation}
\frac{\partial \tilde{F}_{\,n+1}^{n+1}}{\partial y^{n+1}}=-\frac{\partial 
\tilde{F}_{n+1}^{i}}{\partial y^{i}}-\tilde{\Gamma}_{\mu \lambda }^{\mu }%
\tilde{F}_{n+1}^{\lambda }+\tilde{\Gamma}_{\lambda \,n+1}^{n+1}\tilde{F}%
_{n+1}^{\lambda }+\tilde{\Gamma}_{n+1\,n+1}^{i}\tilde{F}_{i}^{n+1}-\tilde{%
\Gamma}_{i\,n+1}^{i}\tilde{F}_{n+1}^{n+1}
\end{equation}
In terms of the components of $\tilde{F}_{n+1}^{n+1}$ and $\tilde{F}%
_{i}^{n+1}$ the equation above may be written as 
\begin{equation}
\frac{\partial \tilde{F}_{n+1}^{n+1}}{\partial y^{n+1}}=-\varepsilon \bar{%
\phi}^{2}\bar{g}^{ij}\frac{\partial \tilde{F}_{i}^{n+1}}{\partial y^{j}}-2%
\tilde{\Gamma}_{i\,n+1}^{i}\tilde{F}_{n+1}^{n+1}+\left( -\varepsilon \frac{%
\partial \left( \bar{\phi}^{2}\bar{g}^{ij}\right) }{\partial y^{j}}%
-\varepsilon \bar{\phi}^{2}\bar{g}^{ij}\tilde{\Gamma}_{k\,j}^{k}+\tilde{%
\Gamma}_{n+1\,n+1}^{i}\right) \tilde{F}_{i}^{n+1}  \label{Fn+1}
\end{equation}

Analogously for $\beta =i$ we obtain 
\begin{equation}
\frac{\partial \tilde{F}_{i}^{n+1}}{\partial y^{n+1}}=\frac{\partial \tilde{F%
}_{n+1}^{n+1}}{\partial y^{i}}+2\tilde{\Gamma}_{n+1\,i}^{n+1}\tilde{F}%
_{n+1}^{n+1}+\left( \tilde{\Gamma}_{n+1\,i}^{k}+\varepsilon \bar{\phi}^{2}%
\bar{g}^{kj}\tilde{\Gamma}_{ij}^{n+1}-\tilde{\Gamma}_{n+1\mu }^{\mu }\delta
_{i}^{k}\right) \tilde{F}_{k}^{n+1}  \label{Fin+1}
\end{equation}
By taking into account (\ref{Fn+1}) and (\ref{Fin+1}) it can be easily shown
by mathematical induction that 
\begin{equation}
\left. \frac{\partial ^{r}\tilde{F}_{\beta }^{n+1}}{\partial \left(
y^{n+1}\right) ^{r}}\right| _{y^{n+1}=0}=0  \label{drF}
\end{equation}
for any integer $r\geq 0.$ As a consequence we conclude that $\tilde{F}%
_{\beta }^{n+1}=0$ in a neighborhood of the origin. Indeed, as $\bar{g}_{ik}$
and $\bar{\phi}$ are analytic at $0\in {\Bbb R}^{n+1}$, then there exists
such a neighborhood in which $\tilde{F}_{\beta }^{n+1}$ can be expressed as
a Taylor series about $0\in {\Bbb R}^{n+1}$, which each term of this series
being null. Since the equation $\tilde{F}_{\beta }^{n+1}=0$ is equivalent to
the equations (\ref{2Rn+1}) and (\ref{Gn+1}), then the lemma is proved.

\subsection{The Cauchy-Kowalewski theorem and the existence of the embedding.%
}

Although essential to the main result to be presented later, Lemma 1 says
nothing about the existence of the solutions $\bar{g}_{ik}$ and $\bar{\phi}.$%
Thus we have to resort to the following theorem.

{\bf Theorem (Cauchy-Kowalewski). }{\it Let us consider the set of partial
differential equations}$:$%
\begin{equation}
\frac{\partial ^{2}u^{A}}{\partial \left( y^{n+1}\right) ^{2}}=F^{A}\left(
y^{\alpha },u^{B},\frac{\partial u^{B}}{\partial y^{\alpha }},\frac{\partial
^{2}u^{B}}{\partial y^{\alpha }\partial y^{i}},\right) ,\qquad A=1,...,m
\label{CK}
\end{equation}
{\it where }$u^{1},..,u^{m}$ {\it are }$m$ {\it unknown functions of the }$%
n+1$ {\it variables }$y^{1},...,y^{n},y^{n+1},$ $\alpha =1,...,n+1,$ $%
i=1,..,n,$ $B=1,...,m.$ {\it \ Also, let }$\xi ^{1},...,\xi ^{m},\eta
^{1},...,\eta ^{m},$ {\it functions of the variables }$y^{1},...,y^{n},$ 
{\it be analytic at }$0\in ${\it \ }${\Bbb R}^{n}.${\it \ If the functions }$%
F^{A}$ {\it are analytic with respect to each of their arguments around the
values evaluated at the point }$y^{1}=...=y^{n}=0,${\it \ then there exists
a unique solution of equations (\ref{CK}) which is analytic at }$0\in {\Bbb R%
}^{n+1}$ {\it and that satisfies the initial condition} 
\begin{eqnarray}
u^{A}\left( y^{1},...,y^{n},0\right) &=&\xi ^{A}\left( y^{1},...,y^{n}\right)
\\
\frac{\partial u^{A}}{\partial y^{n+1}}\left( y^{1},...,y^{n},0\right)
&=&\eta ^{A}\left( y^{1},...,y^{n}\right) ,\qquad A=1,...,m.
\end{eqnarray}

By using (\ref{hik}), we can rewrite (\ref{2Rik}) as 
\begin{eqnarray}
\frac{\partial ^{2}\bar{g}_{ik}}{\partial \left( y^{n+1}\right) ^{2}}
&=&\varepsilon \frac{4\Lambda }{1-n}\bar{\phi}^{2}\bar{g}_{ik}+\frac{1}{\bar{%
\phi}}\frac{\partial \bar{\phi}}{\partial y^{n+1}}\frac{\partial \bar{g}_{ik}%
}{\partial y^{n+1}}-\frac{1}{2}\bar{g}^{jm}\left( \frac{\partial \bar{g}_{ik}%
}{\partial y^{n+1}}\frac{\partial \bar{g}_{jm}}{\partial y^{n+1}}-2\frac{%
\partial \bar{g}_{im}}{\partial y^{n+1}}\frac{\partial \bar{g}_{jk}}{%
\partial y^{n+1}}\right)   \nonumber \\
&&-2\varepsilon \bar{\phi}\left( \frac{\partial ^{2}\bar{\phi}}{\partial
y^{i}\partial y^{k}}-\frac{\partial \bar{\phi}}{\partial y^{j}}\bar{\Gamma}%
_{ik}^{j}\right) -2\varepsilon \bar{\phi}^{2}\bar{R}_{ik}.  \label{eqgik}
\end{eqnarray}
Owing to the symmetry condition $\bar{g}_{ik}=\bar{g}_{ki}$, we can express
Eq. (\ref{eqgik}) in terms of the functions $\bar{g}_{ik}$ with $i\leq k.$
If $\bar{\phi}$ is regarded as a known function, then (\ref{eqgik}) becomes
a set of partial differential equations for the $m=\frac{n\left( n+1\right) 
}{2}$ unknown function $\bar{g}_{ik}\left( i\leq k\right) $. This set of
equations has the same form as (\ref{CK}). We also note that the right-hand
side of (\ref{eqgik}) consists of rational functions of the coordinates $y,$
the functions $\bar{g}_{ik}\left( i\leq k\right) $ and their derivatives (up
to first order with respect to $y^{n+1}$ and up to second order relatively
to the other coordinates). Therefore, if we take $\bar{\phi}$ $\neq 0$
analytic at $0\in {\Bbb R}^{n+1}$ and if the initial conditions 
\begin{eqnarray}
\mathrel{\mathop{\bar{g}_{ik}}\limits_{i\leq k}}%
\left( y^{1},..,y^{n},0\right)  &=&%
\mathrel{\mathop{g_{ik}}\limits_{i\leq k}}%
\left( y^{1},..,y^{n}\right)   \label{cig} \\
\mathrel{\mathop{\frac{\partial \bar{g}_{ik}}{\partial y^{n+1}}}\limits_{i\leq k}}%
\left( y^{1},..,y^{n},0\right)  &=&-2\bar{\phi}\left(
y^{1},..,y^{n},0\right) 
\mathrel{\mathop{\Omega _{ik}}\limits_{i\leq k}}%
\left( y^{1},...,y^{n}\right) ,  \label{cih}
\end{eqnarray}
hold in some neighborhood of the point $0\in $ ${\Bbb R}^{n},$ where $g_{ik}$
and $\Omega _{ik}$ are arbitrary analytic functions with $\left|
g_{ik}\right| \neq 0$ at the origin, then the right-hand side of (\ref{eqgik}%
) will also be analytic at 
\begin{equation}
y^{1}=0...y^{n+1}=0;\left. 
\mathrel{\mathop{\bar{g}_{ik}}\limits_{i\leq k}}%
\right| _{0};\left. 
\mathrel{\mathop{\frac{\partial \bar{g}_{ik}}{\partial y^{1}}}\limits_{i\leq k}}%
\right| _{0}...\left. 
\mathrel{\mathop{\frac{\partial \bar{g}_{ik}}{\partial y^{n+1}}}\limits_{i\leq k}}%
\right| _{0};\left. 
\mathrel{\mathop{\frac{\partial ^{2}\bar{g}_{ik}}{\partial y^{j}\partial y^{m}}}\limits_{i\leq k}}%
\right| _{0}
\end{equation}
We conclude, then, from the Cauchy-Kowalewski theorem, that Eq. (\ref{eqgik}%
) admits a unique solution $\bar{g}_{ik}\left( y^{1},...,y^{n+1}\right) $
which is analytic at $0\in {\Bbb R}^{n+1}$ and also satisfies the given
initial conditions. It should be noted that the determinant $\left| \bar{g}%
_{ik}\right| $ (which due to the initial conditions is non-null at the
origin) remains non null in some open set of ${\Bbb R}^{n+1}$ as a
consequence of the continuity of the solution. These results may be summed
up in the following lemma.

{\bf Lemma 2.}{\it \ Let }$g_{ik}\left( y^{1},...,y^{n}\right) $ {\it and} $%
\Omega _{ik}\left( y^{1},...,y^{n}\right) $, {\it for} $i,k=1,..,n,$ {\it and%
} $\bar{\phi}\left( y^{1},...,y^{n},y^{n+1}\right) ,$ {\it be arbitrary
functions which are analytic at }$0\in ${\it \ }${\Bbb R}^{n}$ {\it and } $%
0\in {\Bbb R}^{n+1},$ {\it respectively, with }$g_{ik}=g_{ki},$ $\left|
g_{ik}\right| \neq 0,$ $\Omega _{ik}=\Omega _{ki}$ {\it in some open set of }%
${\Bbb R}^{n}$ {\it containing }$0\in {\Bbb R}^{n}${\it , and }$\bar{\phi}%
\neq 0$ {\it in some open set of }${\Bbb R}^{n+1}$ {\it containing }$0\in $%
{\it \ }${\Bbb R}^{n+1}${\it . Then there exists a unique set of functions }$%
\bar{g}_{ik}\left( y^{1},...,y^{n},y^{n+1}\right) ,$ {\it which are analytic
at }$0\in ${\it \ }${\Bbb R}^{n+1},${\it \ that satisfy: i) the conditions (%
\ref{ik=ki}), (\ref{det}) and the equation (\ref{2Rik}) in a neighborhood of 
}$0\in $ ${\Bbb R}^{n+1};$ {\it and ii) the initial conditions (\ref{gbar=g}%
) and (\ref{cih}).}

Now if we identify the functions $g_{ik}$ of the initial conditions with the
components of a semi-Riemannian manifold $M^{n},$ then from lemma 1, lemma 2
and theorem 1 we can proof the following theorem:

{\bf Theorem 2. }{\it Let }$M^{n}$ {\it be a n-dimensional semi-Riemannian
manifold with metric given by} 
\[
ds^{2}=g_{ik}dx^{i}dx^{j},
\]
{\it in coordinate system }$\left\{ x^{i}\right\} $ {\it of }$M^{n}.$ {\it %
Let }$p\in M^{n}$ {\it have coordinates }$x_{p}^{1}=...=x_{p}^{n}=0.$ {\it %
Then, }$M^{n}$ {\it has a local isometric and analytic embedding (at the
point }$p$) {\it in a (n+1)-dimensional Einstein space with cosmological
constant }$\Lambda $ {\it if and only if there exist functions }$\Omega
_{ik}\left( x^{1},...,x^{n}\right) ,${\it \ (}$i,k=1,..,n),$ {\it that are
analytic at }$0\in {\Bbb R}^{n}$ {\it and such that } 
\begin{eqnarray}
\Omega _{ik} &=&\Omega _{ki}  \label{eq1} \\
g^{jk}\left( \nabla _{j}\Omega _{ik}-\nabla _{i}\Omega _{jk}\right)  &=&0
\label{eq2} \\
g^{ik}g^{jm}\left( R_{ijkm}+\varepsilon \left( \Omega _{ik}\Omega
_{jm}-\Omega _{jk}\Omega _{im}\right) \right)  &=&-2\Lambda .  \label{eq3}
\end{eqnarray}

{\it Proof. }Let ${\cal M}_{\Lambda }^{n+1}$ be the collection of all $%
\left( n+1\right) -$dimensional Einstein spaces with cosmological constant $%
\Lambda .$ If $M^{n}$ has a local and analytic embedding in ${\cal M}%
_{\Lambda }^{n+1},$ at the point $p,$ then in accordance with theorem 1,
there exist functions $\bar{g}_{ik}\left( x^{1},...,x^{n+1}\right) ,$
satisfying (\ref{gbar=g}), and $\bar{\phi}\left( x^{1},...,x^{n+1}\right) $
that are analytic at $0\in {\Bbb R}^{n+1}$ such that 
\begin{equation}
ds^{2}=\bar{g}_{ik}dx^{i}dx^{k}+\varepsilon \bar{\phi}^{2}dx^{n+1}dx^{n+1}
\end{equation}
is the line element of some member of ${\cal M}_{\Lambda }^{n+1}$ expressed
in a conveniently chosen coordinate system. Therefore , this metric
satisfies the equations (\ref{2Rik}), (\ref{2Rn+1}) and (\ref{Gn+1}) in a
neighborhood of $0\in $ ${\Bbb R}^{n+1}$. In particular, this is true for
points lying on the hypersurface $x^{n+1}=0,$ where $\bar{g}_{ik}=g_{ik},$
from (\ref{gbar=g}){\it .} Thus, if we define $\Omega _{ik}(x^{1},...,x^{n})=%
\bar{\Omega}_{ik}(x^{1},...,x^{n},0),$ then, the functions $\Omega _{ik}$
necessarily satisfy (\ref{eq1}), (\ref{eq2}) and (\ref{eq3}).

Let us consider the sufficient condition. First, we choose $\bar{\phi}\left(
x^{1},...,x^{n+1}\right) \neq 0$ and analytic at $0\in {\Bbb R}^{n+1}.$
According to lemma 2, there exists a unique set of functions $\bar{g}%
_{ik}\left( x^{1},...,x^{n+1}\right) $ satisfying (\ref{gbar=g}), (\ref
{ik=ki}), (\ref{det}), (\ref{hik}), (\ref{2Rik}) and the condition $\bar{%
\Omega}_{ik}\left( x^{1},...,x^{n},0\right) =\Omega _{ik}\left(
x^{1},...,x^{n}\right) .$ If $\Omega _{ik}$ and $g_{ik}$ satisfy (\ref{eq2})
and (\ref{eq3}), then, from lemma 1 the functions $\bar{g}_{ik}\left(
x^{1},...,x^{n+1}\right) $ satisfy (\ref{2Rik}), (\ref{2Rn+1}) and (\ref
{Gn+1}) in a neighborhood of $0\in $ ${\Bbb R}^{n+1},$ which in turn implies
that the line element formed with $\bar{g}_{ik}$ and $\bar{\phi}$ is that of
an Einstein space with cosmological constant $\Lambda .$ Then, theorem 1
tell us that $M^{n}$ has a local isometric and analytical embedding in $%
{\cal M}_{\Lambda }^{n+1}.\square $

We now want to show that once the functions $g_{ik}$ are given, the
equations (\ref{eq1}), (\ref{eq2}) and (\ref{eq3}) always admit a solution
for $\Omega _{ik}.$ These equations constitute a set of $n$ partial
differential equations (Eq. (\ref{eq2})) plus a constraint equation (Eq.(\ref
{eq3})) for $\frac{n\left( n+1\right) }{2}$ independent functions $\Omega
_{ik}$. Except for $n=1,$ the number of unknown functions is greater than
(or equal to ($n=2$)) the number of equations. Out of the set of functions $%
\Omega _{ik}$ we pick $n$ functions to be regarded as the unknowns. We
proceed to put (\ref{eq2}) in the form required for application of the
Cauchy-Kowalewski (first-order derivative version) theorem to assure the
existence of the solution. The detailed proof is a bit laborious, so we
shall omit some of its parts.

For the sake of the argument and with no loss of generality we assume that
we are using a coordinate system in which $g_{11}\neq 0$ and $g_{1k}=0,$ $%
k=2,...,n.$ Thus, $g^{11}=\frac{1}{g_{11}}$ and $g^{1k}=0.$ Eq. (\ref{eq2})
can be written as 
\begin{equation}
g^{rs}\left( \Omega _{sk,r}-\Omega _{rs,k}+\Omega _{rt}\Gamma
_{sk}^{t}-\Omega _{kt}\Gamma _{sr}^{t}\right) =0.  \label{eq2.1}
\end{equation}
Recalling that $\Omega _{ik}=\Omega _{ki}$, it is not difficult to see that (%
\ref{eq2.1}) may be put in the form 
\begin{equation}
g^{rs}\left( 
\mathrel{\mathop{\Omega _{sk,r}}\limits_{s\leq k}}%
+%
\mathrel{\mathop{\Omega _{ks,r}}\limits_{k<s}}%
-2%
\mathrel{\mathop{\Omega _{rs,k}}\limits_{r<s}}%
\right) -g^{rr}\Omega _{rr,k}+g^{rs}\left( 
\mathrel{\mathop{\Omega _{tr}}\limits_{t\leq r}}%
\Gamma _{sk}^{t}+%
\mathrel{\mathop{\Omega _{rt}}\limits_{r<t}}%
\Gamma _{sk}^{t}-%
\mathrel{\mathop{\Omega _{tk}}\limits_{t\leq k}}%
\Gamma _{sr}^{t}-%
\mathrel{\mathop{\Omega _{kt}}\limits_{k<t}}%
\Gamma _{sr}^{t}\right) =0.  \label{78}
\end{equation}

Likewise and taking advantage of the special form of the metric we can write
the Eq. (\ref{eq3}) as 
\begin{eqnarray}
2g^{11}\Omega _{11}%
\mathrel{\mathop{g^{rs}}\limits_{r,s>1}}%
\left( 
\mathrel{\mathop{\Omega _{rs}}\limits_{r\leq s}}%
+%
\mathrel{\mathop{\Omega _{sr}}\limits_{s<r}}%
\right) -2g^{11}%
\mathrel{\mathop{g^{rs}}\limits_{r,s>1}}%
\Omega _{1r}\Omega _{1s} &&+  \label{79} \\
\mathrel{\mathop{+g^{rs}g^{tu}}\limits_{r,s,t,u>1}}%
\left[ \left( 
\mathrel{\mathop{\Omega _{rs}}\limits_{r\leq s}}%
+%
\mathrel{\mathop{\Omega _{sr}}\limits_{s<r}}%
\right) \left( 
\mathrel{\mathop{\Omega _{tu}}\limits_{t\leq u}}%
+%
\mathrel{\mathop{\Omega _{ut}}\limits_{u<t}}%
\right) -\left( 
\mathrel{\mathop{\Omega _{ru}}\limits_{r\leq u}}%
+%
\mathrel{\mathop{\Omega _{ur}}\limits_{u<r}}%
\right) \left( 
\mathrel{\mathop{\Omega _{st}}\limits_{s\leq t}}%
+%
\mathrel{\mathop{\Omega _{ts}}\limits_{t<s}}%
\right) \right] +\varepsilon R &=&-\varepsilon \Lambda  \nonumber
\end{eqnarray}

Our next step is to identify in the Eq. (\ref{78}) the terms containing
derivatives of $\Omega _{ik}$ with respect to the coordinate $x^{1}.$ Let us
consider first the case $k=1.$ Thus, we have 
\begin{equation}
g^{rs}\left( 
\mathrel{\mathop{\Omega _{1s,r}}\limits_{r,s>1}}%
-2%
\mathrel{\mathop{\Omega _{rs,1}}\limits_{1<r<s}}%
\right) -g^{rr}%
\mathrel{\mathop{\Omega _{rr,1}}\limits_{r>1}}%
+%
\mathrel{\mathop{g^{rs}}\limits_{r,s>1}}%
\left( 
\mathrel{\mathop{\Omega _{tr}}\limits_{t\leq r}}%
\Gamma _{s1}^{t}+%
\mathrel{\mathop{\Omega _{rt}}\limits_{r<t}}%
\Gamma _{s1}^{t}-\Omega _{11}\Gamma _{sr}^{1}-%
\mathrel{\mathop{\Omega _{1t}}\limits_{t<1}}%
\Gamma _{sr}^{t}\right) =0  \label{80}
\end{equation}
In order to write (\ref{80}) in the form specified by Cauchy-Kowalewski (in
its first-derivative version) we should decide what among the functions $%
\Omega _{ik}$ $\left( i\leq k\right) $ are to be chosen as unknowns. We also
note that since $\left| g_{ik}\right| \neq 0$ there exists at least an index 
$r^{\prime }>1$ such that $g^{r^{\prime }n}\neq 0$. We choose $\Omega
_{r^{\prime }n}$ as one of the unknown functions and solve (\ref{80}) for $%
\frac{\partial \Omega _{r^{\prime }n}}{\partial x^{1}}.$ Thus, it is
possible to put (\ref{80}) in the form 
\begin{align}
\frac{\partial \Omega _{r^{\prime }n}}{\partial x^{1}}& =\frac{1}{%
g^{r^{\prime }n}\left( \delta _{r^{\prime }n}-2\right) }\left[ -%
\mathrel{\mathop{g^{rs}}\limits_{r,s>1}}%
\Omega _{1s,r}+2%
\mathrel{\mathop{g^{rs}\Omega _{rs,1}}\limits_{%
{\textstyle{1<r<s \atopwithdelims.. r,s\neq r^{\prime },n}}}}%
+g^{rr}%
\mathrel{\mathop{\Omega _{rr,1}}\limits_{%
{\textstyle{r>1 \atopwithdelims.. r\neq r^{\prime }}}}}%
+\right.  \label{81} \\
& \left. +g^{r^{\prime }r^{\prime }}\Omega _{r^{\prime }r^{\prime },1}\left(
1-\delta _{r^{\prime }n}\right) -%
\mathrel{\mathop{g^{rs}}\limits_{r,s>1}}%
\left( 
\mathrel{\mathop{\Omega _{tr}}\limits_{t\leq r}}%
\Gamma _{s1}^{t}+%
\mathrel{\mathop{\Omega _{rt}}\limits_{r<t}}%
\Gamma _{s1}^{t}-\Omega _{11}\Gamma _{sr}^{1}-%
\mathrel{\mathop{\Omega _{1t}}\limits_{t<1}}%
\Gamma _{sr}^{t}\right) \right] ,  \nonumber
\end{align}
where no sum over $r^{\prime }$ is implied.

For $k\geq 2$ we have 
\begin{eqnarray}
\frac{\partial \Omega _{1k}}{\partial x^{1}} &=&g_{11}\left[ -%
\mathrel{\mathop{g^{rs}}\limits_{r,s>1}}%
\left( 
\mathrel{\mathop{\Omega _{sk,r}}\limits_{s\leq k}}%
+%
\mathrel{\mathop{\Omega _{ks,r}}\limits_{k<s}}%
-2%
\mathrel{\mathop{\Omega _{rs,k}}\limits_{r<s}}%
\right) -g^{11}\Omega _{11,k}-g^{rr}%
\mathrel{\mathop{\Omega _{rr,k}}\limits_{r>1}}%
\right.  \label{82} \\
&&\left. -g^{rs}\left( 
\mathrel{\mathop{\Omega _{tr}}\limits_{t\leq r}}%
\Gamma _{sk}^{t}+%
\mathrel{\mathop{\Omega _{rt}}\limits_{r<t}}%
\Gamma _{sk}^{t}-%
\mathrel{\mathop{\Omega _{tk}}\limits_{t\leq k}}%
\Gamma _{sr}^{t}-%
\mathrel{\mathop{\Omega _{kt}}\limits_{k<t}}%
\Gamma _{sr}^{t}\right) \right] ,\qquad k\geq 2.  \nonumber
\end{eqnarray}

From (\ref{79}) we can express $\Omega _{11}$ in terms of the other $\Omega
_{ik}.$Thus 
\begin{eqnarray}
\Omega _{11} &=&\frac{1}{2g^{11}%
\mathrel{\mathop{g^{rs}}\limits_{r,s>1}}%
\left( 
\mathrel{\mathop{\Omega _{rs}}\limits_{r\leq s}}%
+%
\mathrel{\mathop{\Omega _{sr}}\limits_{s<r}}%
\right) }\left[ 2g^{11}%
\mathrel{\mathop{g^{rs}}\limits_{r,s>1}}%
\Omega _{1r}\Omega _{1s}\right.  \label{83} \\
&&\left. -%
\mathrel{\mathop{g^{rs}g^{tu}}\limits_{r,s,t,u>1}}%
\left[ \left( 
\mathrel{\mathop{\Omega _{rs}}\limits_{r\leq s}}%
+%
\mathrel{\mathop{\Omega _{sr}}\limits_{s<r}}%
\right) \left( 
\mathrel{\mathop{\Omega _{tu}}\limits_{t\leq u}}%
+%
\mathrel{\mathop{\Omega _{ut}}\limits_{u<t}}%
\right) -\left( 
\mathrel{\mathop{\Omega _{ru}}\limits_{r\leq u}}%
+%
\mathrel{\mathop{\Omega _{ur}}\limits_{u<r}}%
\right) \left( 
\mathrel{\mathop{\Omega _{st}}\limits_{s\leq t}}%
+%
\mathrel{\mathop{\Omega _{ts}}\limits_{t<s}}%
\right) \right] -\varepsilon \left( R+\Lambda \right) \right]  \nonumber
\end{eqnarray}

Finally, substituting $\Omega _{11}$ from (\ref{83}) into (\ref{81}) and (%
\ref{82}) we obtain a set of partial differential equations for the
functions $\Omega _{ik}.$ If we regard the functions $\Omega _{ik}$ $\left(
i\leq k\right) $ with $i>1$ and $\left( i,k\right) \neq \left( r^{\prime
},n\right) $ as analytic functions already known, then we apply the
Cauchy-Kowalewski theorem (first derivative version) to this set of
differential equations considering $\Omega _{1k}\left( k>1\right) $ and $%
\Omega _{r^{\prime }n}$ as the unknown functions. We, then, choose $\Omega
_{ik}\left( x^{1},..,x^{n}\right) $ $\left( i\leq k,i>1,\left( i,k\right)
\neq \left( r^{\prime },n\right) \right) $ and the initial conditions $%
\Omega _{1k}\left( 0,x^{2},...,x^{n}\right) =f_{k}\left(
x^{2},...,x^{n}\right) $ $\left( k>1\right) $ and $\Omega _{r^{\prime
}n}\left( 0,x^{2},...,x^{n}\right) =f_{1}\left( x^{2},...,x^{n}\right) .$ Of
course the chosen functions must be analytic at $0\in {\Bbb R}^{n}$ and
satisfy the condition 
\begin{equation}
\left. 
\mathrel{\mathop{g^{rs}}\limits_{r,s>1}}%
\left( 
\mathrel{\mathop{\Omega _{rs}}\limits_{r\leq s}}%
+%
\mathrel{\mathop{\Omega _{sr}}\limits_{s<r}}%
\right) \right| _{0}\neq 0.  \label{84}
\end{equation}
It should be noted that it is always possible to have functions $\Omega
_{ik} $ that satisfy the condition above. For instance, if we take $\Omega
_{ik}=0$ $\left( i\leq k,i>1,\left( i,k\right) \neq \left( r^{\prime
},n\right) \right) $ this condition reduces to $\left. g^{r^{\prime
}n}\Omega _{r^{\prime }n}\right| _{0}\neq 0.$ Hence we just choose $\Omega
_{r^{\prime }n}\neq 0$.

Once we have specified the functions $\Omega _{ik}\left(
x^{1},..,x^{n}\right) $ $\left( i\leq k,i>1,\left( i,k\right) \neq \left(
r^{\prime },n\right) \right) $ the right-hand side of (\ref{81}) and (\ref
{82}) becomes function of the arguments 
\begin{equation}
x^{1},...,x^{n};%
\mathrel{\mathop{\Omega _{1k}}\limits_{k>1}}%
,\Omega _{r^{\prime }n};%
\mathrel{\mathop{\Omega _{1k,j}}\limits_{k,j>1}}%
,%
\mathrel{\mathop{\Omega _{r^{\prime }n,j}}\limits_{j>1}}%
\end{equation}
which is analytic at 
\begin{equation}
x^{1}=0,...,x^{n}=0;\left. 
\mathrel{\mathop{\Omega _{1k}}\limits_{k>1}}%
\right| _{0},\left. \Omega _{r^{\prime }n}\right| _{0};\left. 
\mathrel{\mathop{\Omega _{1k,j}}\limits_{k,j>1}}%
\right| _{0},\left. 
\mathrel{\mathop{\Omega _{r^{\prime }n,j}}\limits_{j>1}}%
\right| _{0}.
\end{equation}
Therefore, the Cauchy-Kowalewski theorem asserts that there exists a unique
set of functions $\Omega _{1k}\left( x^{1},x^{2},...,x^{n}\right) $ $\left(
k>1\right) $ and $\Omega _{r^{\prime }n}\left( x^{1},x^{2},...,x^{n}\right) ,
$ analytic at $0\in {\Bbb R}^{n+1}$ which satisfy the equations (\ref{81}) e
(\ref{82}). We determine $\Omega _{11}$ from (\ref{83}) by taking the chosen
functions $\Omega _{ik}\left( x^{1},..,x^{n}\right) $ $\left( i\leq
k,i>1,\left( i,k\right) \neq \left( r^{\prime },n\right) \right) $ and the
solutions of the system of equations. From (\ref{84}) and due to the
analyticity of $g_{ik}$ and of the solution we conclude that $\Omega _{11}$
is analytic at the origin. Therefore, the existence of analytic functions $%
\Omega _{ik}$ satisfying (\ref{eq1}), (\ref{eq2}) and (\ref{eq3}) is
demonstrated. The above may be summarized in the following lemma:

{\bf Lemma 3}. {\it Let }$g_{ik}\left( x^{1},...,x^{n}\right) $ {\it and }$%
\Omega _{ik}\left( x^{1},...,x^{n}\right) $ $\left( i\leq k,i>1,\left(
i,k\right) \neq \left( r^{\prime },n\right) \right) $ {\it be analytic
functions at the origin }$0\in {\Bbb R}^{n},$ {\it \ with }$\Omega _{ik}$ 
{\it satisfying the initial conditions }$\Omega _{1k}\left(
0,x^{2},...,x^{n}\right) =f_{k}\left( x^{2},...,x^{n}\right) $ $\left(
k>1\right) $ and $\Omega _{r^{\prime }n}\left( 0,x^{2},...,x^{n}\right)
=f_{1}\left( x^{2},...,x^{n}\right) ,${\it \ where }$f_{k}$ {\it are
analytic at }$0\in {\Bbb R}^{n}.$ {\it \ If, in addition, the condition (\ref
{84}) is fulfilled then there exists a unique set of functions }$\Omega
_{ik}\left( x^{1},...,x^{n}\right) ${\it \ }$\left( i,k=1,...,n\right) ,$%
{\it \ analytic at }$0\in {\Bbb R}^{n-1}${\it , which satisfy the equations (%
\ref{eq1}), (\ref{eq2}) e (\ref{eq3}).}

Therefore, according to the lemma above, if we are given a set of analytic
functions $g_{ik}$, then the existence of analytic functions $\Omega _{ik}$
which satisfy the equations (\ref{eq1}), (\ref{eq2}) and (\ref{eq3}) is
assured. In this way lemma 3 tell us that the sufficient conditions of
theorem (2) are satisfied, so we can to state the final theorem.

{\bf Theorem 3.} {\bf \ }{\it Let }$M^{n}$ $(n>1)$ {\it be a semi-Riemannian
space with line element} 
\[
ds^{2}=g_{ik}dx^{i}dx^{k}, 
\]
{\it expressed in a coordinate system which covers a neighborhood of a point 
}$p\in M^{n}$ {\it whose coordinates are }$x_{p}^{1}=...=x_{p}^{n}=0.$ {\it %
If }$g_{ik}$ {\it are analytic functions at }$0\in {\Bbb R}^{n},${\it \ then 
}$M^{n}${\it \ can be embedded at }$p$ {\it in some }$\left( n+1\right) $%
{\it -dimensional Einstein space }$N^{n+1}\in {\cal M}_{\Lambda }.$

Two comments are in order. First, if the $\frac{n\left( n-1\right) }{2}-1$
specified arbitrary functions $\Omega _{ik}\left( x^{1},...,x^{n}\right) $ $%
\left( i\leq k,i>1,\left( i,k\right) \neq \left( r^{\prime },n\right)
\right) $ obey the conditions

i) the functions $\Omega _{ik}$ $\left( i\leq k,i>1,\left( i,k\right) \neq
\left( r^{\prime },n\right) \right) $ are analytic at $0\in {\Bbb R}^{n};$

ii) the $n$ functions $\Omega _{1k}\left( 0,x^{2},...,x^{n}\right)
=f_{k}\left( x^{2},...,x^{n}\right) $ $\left( k>1\right) $ and $\Omega
_{r^{\prime }n}\left( 0,x^{2},...,x^{n}\right) =f_{1}\left(
x^{2},...,x^{n}\right) $ are analytic at $0\in {\Bbb R}^{n-1},$ with $\left. 
\mathrel{\mathop{g^{rs}}\limits_{r,s>1}}%
\left( 
\mathrel{\mathop{\Omega _{rs}}\limits_{r\leq s}}%
+%
\mathrel{\mathop{\Omega _{sr}}\limits_{s<r}}%
\right) \right| _{0}\neq 0;$

iii) a function $\bar{\phi}\left( x^{1},...,x^{n+1}\right) \neq 0$, analytic
at $0\in {\Bbb R}^{n+1},$ is chosen;

then the line element of the embedding space as referred to in theorem 1 is
unique.$.$

Second, if we consider the case $\Lambda =0,$ then clearly this theorem
reduces to Campbell-Magaard theorem, which establishes the existence of
local analytic embedding of any Riemannian manifold in the set of Ricci-flat
spaces. In this sense, theorem (3) is a generalization of the
Campbell-Magaard theorem.

\section{Applications of the extended Campbell-Magaard theorem}

Let us consider some cases where $\left( M^{n},g\right) $ is a Lorentzian
manifold of dimension $n=4.$ We know from theorem (3) that there exists at
least one Einstein space of dimension $n=5$ in which $\left( M^{4},g\right) $
can be embedded. In this section, we shall discuss the embedding of
Minkowski and Schwarzschild space-time and exhibit explicitly the
five-dimensional embedding Einstein spaces.

It is interesting to observe that the very proof of theorem (3) suggests a
procedure for obtaining the embedding. Indeed, let $g_{ik}$ be the
components of the metric of $M^{n}$ in a coordinate system and let $g_{ik}$
be analytic at the origin of ${\Bbb R}^{n}.$ We first take 
\begin{equation}
\bar{g}_{ik}\left( x^{1},..,x^{n},0\right) =g_{ik}\left(
x^{1},..,x^{n}\right) ,
\end{equation}
and then look for functions $\Omega _{ik}$ which satisfy the equations (\ref
{eq1}), (\ref{eq2}) and (\ref{eq3}) at $\Sigma _{0}.$ Given that the
function $\bar{\phi}$ is arbitrary we can take $\bar{\phi}=1$ for
simplicity. Then, the derivative of $\bar{g}_{ik}$ with respect to the extra
coordinate at $\Sigma _{0}$ will be given by 
\begin{equation}
\frac{\partial \bar{g}_{ik}}{\partial x^{n+1}}\left( x^{1},..,x^{n},0\right)
=-2\Omega _{ik}\left( x^{1},...,x^{n}\right) .
\end{equation}
From $\bar{g}_{ik}$ and the derivative $\frac{\partial \bar{g}_{ik}}{%
\partial x^{n+1}}$ we now calculate $\frac{\partial ^{2}\bar{g}_{ik}}{%
\partial \left( x^{n+1}\right) ^{2}}$ at $\Sigma _{0}$ by using (\ref{eqgik}%
): 
\begin{equation}
\frac{\partial ^{2}\bar{g}_{ik}}{\partial \left( x^{n+1}\right) ^{2}}\left(
x^{1},..,x^{n},0\right) =\varepsilon \frac{4\Lambda }{1-n}%
g_{ik}-2g^{jm}\left( \Omega _{ik}\Omega _{jm}-2\Omega _{im}\Omega
_{jk}\right) -2\varepsilon R_{ik}.
\end{equation}
The equation (\ref{eqgik}) is supposed to hold in an open set of ${\Bbb R}%
^{n+1},$ hence it can be differentiated with respect to $x^{n+1}.$ Doing
this, we are able to put the third derivative of $\bar{g}_{ik}$ as a
function of terms which with respect to the coordinate $x^{n+1}$ contain at
most second derivatives. In this manner we can obtain $\frac{\partial ^{3}%
\bar{g}_{ik}}{\partial \left( x^{n+1}\right) ^{3}}\left(
x^{1},..,x^{n},0\right) $ from the derivatives $\frac{\partial \bar{g}_{ik}}{%
\partial x^{n+1}}\left( x^{1},..,x^{n},0\right) $ and $\frac{\partial ^{2}%
\bar{g}_{ik}}{\partial \left( x^{n+1}\right) ^{2}}\left(
x^{1},..,x^{n},0\right) $ already calculated. This process can go on
indefinitely by sucessively differentiating Eq. (\ref{eqgik}) with respect $%
x^{n+1}$ and such a procedure will give us the derivatives of all orders
calculated at $\Sigma _{0}.$ By collecting all these terms we obtain the
expression of $\bar{g}_{ik}$ into a power series about the origin $0\in 
{\Bbb R}^{n+1},$which by the Cauchy-Kowalewski theorem, is convergent in a
neighborhood of $0\in $ ${\Bbb R}^{n+1}.$ Thus, the functions $\bar{g}_{ik}$
and $\bar{\phi}=1$ together constitute the components of the sought
embedding Einstein space.

Let us apply these ideas by first considering the embedding of Minkowski
space-time. In Cartesian coordinates the metric of Minkowski spacetime has
the form 
\begin{equation}
ds^{2}=-dt^{2}+dx^{2}+dy^{2}+dz^{2},
\end{equation}
hence the components $g_{ik}=\eta _{ik}=diag\left( -1,1,1,1\right) $ are
analytic, and, by theorem (3), it can be embedded in a five-dimensional
Einstein space with arbitrary cosmological constant $\Lambda .$ The first
task is to find functions $\Omega _{ik}$ which satisfy the equations (\ref
{eq1}), (\ref{eq2}) and (\ref{eq3}). As Minkowski spacetime has no curvature
the equation (\ref{eq3}) yields 
\begin{equation}
\eta ^{ik}\eta ^{jm}\left( \Omega _{ik}\Omega _{jm}-\Omega _{jk}\Omega
_{im}\right) =-2\varepsilon \Lambda ,
\end{equation}

It is easy to verify that the choice $\Omega _{ik}=-\frac{c}{2}\eta _{ik},$
where $c$ is a constant, solves the above equation if we take $c=\sqrt{-%
\frac{2\varepsilon \Lambda }{3}}.$ Also, this choice satisfies (\ref{eq2}).
On the other hand, choosing $\bar{\phi}=1,$ equation (\ref{cih}) becomes 
\begin{equation}
\left. \frac{\partial \bar{g}_{ik}}{\partial u}\right| _{u=0}=c\eta _{ik}.
\end{equation}
Then, from (\ref{eqgik}) we obtain 
\begin{eqnarray}
\left. \frac{\partial ^{2}\bar{g}_{ik}}{\partial u^{2}}\right| _{u=0}
&=&\left. \left[ \varepsilon \frac{4\Lambda }{1-4}\bar{g}_{ik}-\frac{1}{2}%
\bar{g}^{jm}\left( \frac{\partial \bar{g}_{ik}}{\partial u}\frac{\partial 
\bar{g}_{jm}}{\partial u}-2\frac{\partial \bar{g}_{im}}{\partial u}\frac{%
\partial \bar{g}_{jk}}{\partial u}\right) \right] \right| _{u=0}=  \nonumber
\\
&=&-\varepsilon \frac{4\Lambda }{3}\eta _{ik}-\frac{c^{2}}{2}\eta
^{jm}\left( \eta _{ik}\eta _{jm}-2\eta _{im}\eta _{jk}\right) =c^{2}\eta
_{ik}.
\end{eqnarray}

Following the procedure outlined formerly the higher-order derivatives can
be easily calculated and we finally obtain 
\begin{equation}
\bar{g}_{ik}=\sum_{p=0}^{\infty }\frac{c^{p}}{p!}u^{p}\eta _{ik}=e^{cu}\eta
_{ik}
\end{equation}
Therefore, Minkowski spacetime can be embedded in the Einstein space whose
metric is given by 
\begin{equation}
ds^{2}=e^{\sqrt{-\frac{2\varepsilon \Lambda }{3}}u}\left(
-dt^{2}+dx^{2}+dy^{2}+dz^{2}\right) +\varepsilon du^{2}.
\end{equation}
and $u=0$ corresponds to the sought embedding. Of course the choice of $%
\varepsilon $ depends upon the given $\Lambda $, since we must have $%
\varepsilon \Lambda <0.$ If want $\varepsilon $ and $\Lambda $ to have the
same sign, then another choice of $\Omega _{ik}$ has to be made. If $\Lambda 
$ is negative, then the embedding space is closely related to the so-called
bulk, in the Randall-Sundrum braneworld scenario \cite{randall}.

As a second example let us consider the embedding of Schwarzschild spacetime
whose geometry can be described by the line element 
\begin{equation}
ds^{2}=-\left( 1-\frac{2m}{r}\right) dt^{2}+\left( 1-\frac{2m}{r}\right)
^{-1}dr^{2}+r^{2}d\theta ^{2}+r^{2}sen^{2}\theta d\varphi ^{2},
\end{equation}
where $m$ is a constant. Except for $r=0$ and $r=2m$ the metric components
are analytic in the subset of ${\Bbb R}^{4}$ corresponding to the range of
the coordinates. Though theorem (3) seems to single out the point $\left(
0,0,0,0\right) $ as far as the embedding is concerned it should be clear
that there is nothing special about this point and the embedding can also be
achieved at any other point where the metric components are analytic.

As Schwarzschild spacetime is Ricci-flat the equation (\ref{eq3}) yields 
\begin{equation}
g^{ik}g^{jm}\left( \Omega _{ik}\Omega _{jm}-\Omega _{jk}\Omega _{im}\right)
=-2\varepsilon \Lambda .
\end{equation}
A solution of this equation is given by 
\begin{equation}
\Omega _{ik}=-\frac{c}{2}g_{ik},
\end{equation}
where $g_{ik}$ are the metric components of Schwarzschild spacetime and $c=%
\sqrt{-\frac{2\varepsilon \Lambda }{3}}$ . Due to the compatibility
condition $\nabla _{j}g_{ik}=0,$ it is immediately seen that the equation (%
\ref{eq2}) is satisfied. Again, let us choose $\bar{\phi}=1.$ At this point
we could repeat the iterative procedure employed in the previous example and
obtain the metric components as a power series. Though this method is quite
direct, sometimes it becomes a bit laborious. Here, let us follow a
different route. We shall assume the {\it ansatz} 
\begin{equation}
\bar{\Omega}_{ik}\left( t,r,\theta ,\varphi ,u\right) =-\frac{c\left(
u\right) }{2}g_{ik}\left( t,r,\theta ,\varphi \right) .
\end{equation}
For $u=0$ the equations (\ref{eq2}) e (\ref{eq3}) will hold provided that $%
c\left( 0\right) =\sqrt{-\frac{2\varepsilon \Lambda }{3}}$ . Now from the
definition of $\bar{\Omega}_{ik},$ with $\bar{\phi}=1,$we have 
\begin{equation}
\frac{\partial \bar{g}_{ik}}{\partial u}=c\left( u\right) g_{ik}.
\end{equation}
After integrating the equation above and taking into account the initial
conditions imposed on $\bar{g}_{ik}$ we have $\bar{g}_{ik}\left( t,r,\theta
,\varphi ,u\right) =f\left( u\right) g_{ik},$ where $f\left( u\right) =\int
c\left( w\right) dw$, with $f\left( 0\right) =1.$ On the other hand the
Ricci tensor $\bar{R}_{ik}$ associated with $\bar{g}_{ik}$ vanishes
everywhere, so (\ref{eqgik}) yields the ordinary differential equation for $%
f\left( u\right) :$%
\begin{equation}
f^{\prime \prime }+\frac{f^{\prime 2}}{f}+\frac{4\varepsilon \Lambda }{3}f=0.
\label{f}
\end{equation}
The solution of (\ref{f}) is given by 
\begin{equation}
f\left( u\right) =e^{\sqrt{-\frac{2\varepsilon \Lambda }{3}}u}.
\end{equation}
Hence the conditions $c\left( 0\right) =\sqrt{-\frac{2\varepsilon \Lambda }{3%
}}$ and $f\left( 0\right) =1$ are satisfied. We conclude then that
Schwarzschild spacetime can be embedded in the Einstein space: 
\begin{equation}
ds^{2}=e^{\sqrt{-\frac{2\varepsilon \Lambda }{3}}u}\left[ -\left( 1-\frac{2m%
}{r}\right) dt^{2}+\left( 1-\frac{2m}{r}\right) ^{-1}dr^{2}+r^{2}d\theta
^{2}+r^{2}sen^{2}\theta d\varphi ^{2}\right] +\varepsilon du^{2},
\end{equation}
the embedding taking place for $u=0.$

A third application of theorem 3 may be illustrated by turning our attention
to a more general situation. Suppose $\left( M^{4},g\right) $ is an Einstein
space, that is 
\begin{equation}
R_{ik}=-\lambda g_{ik}.
\end{equation}
Our aim is to find a five-dimensional $\left( M^{5},\tilde{g}\right) $ with
a given arbitrary cosmological constant $\Lambda $ in which $\left(
M^{4},g\right) $ can be embedded.

As in the former example let us assume that 
\begin{equation}
\bar{g}_{ik}=f\left( u\right) g_{ik}.
\end{equation}
with $f\left( 0\right) =1$. Since we now have $R=-4\lambda ,$ the equation (%
\ref{eq3}) can be satisfied only if $f^{\prime }\left( 0\right) =$ $\sqrt{-%
\frac{2\varepsilon \Lambda }{3}+\frac{4\varepsilon \lambda }{3}}.$

Owing to the fact that $\bar{g}_{ik}$ and $g_{ik}$ are related by a
conformal transformation which depends on the extra coordinate $u$ only, it
follows that $\bar{R}_{ik}=R_{ik}=-\lambda g_{ik}$. Thus, Eq. (\ref{eqgik})
is equivalent to 
\begin{equation}
f^{\prime \prime }+\frac{f^{\prime 2}}{f}+\frac{4\varepsilon \Lambda }{3}%
f-2\varepsilon \lambda =0.
\end{equation}
It is not difficult to show that the solution which satisfies the initial
conditions imposed on $f$ is given by \cite{lidsey} 
\begin{equation}
f(u)=\left[ \cosh \left( \sqrt{-\frac{\varepsilon \Lambda }{6}}u\right)
+\left( 1-2\frac{\lambda }{\Lambda }\right) \sinh \left( \sqrt{-\frac{%
\varepsilon \Lambda }{6}}u\right) \right] ^{2}.  \label{f1}
\end{equation}
In this way we conclude that the Einstein space $\left( M^{4},g\right) $ can
be embedded in the Einstein space 
\begin{equation}
ds^{2}=f(u)g_{ik}dx^{i}dx^{j}+\varepsilon du^{2},
\end{equation}
with $f\left( u\right) $ being given by (\ref{f1}).

\section{Conclusion}

The recent appearance of physical models which regard the ordinary spacetime
as a hypersurface embedded in a five-dimensional manifold has naturally
raised the question of what kind of mathematical conditions both the
embedded and the spaces are subject to. An answer to this question
necessarily involves a careful account of the mathematical theory of
embedding. Particularly useful and clarifying are the Campbell-Magaard
theorem and its extension to the case in which the embedding manifold is an
Einstein space. Belonging to the latter kind is the embedding considered in
the Randall-Sundrum braneworld. On the other hand embeddings in Ricci-flat
five-dimensional manifolds are crucial for the non-compactified approach to
Kaluza-Klein gravity \cite{wesson}. Of course, if according to some new
physical model the five-dimensional surrounding manifolds should obey some
field equations, e.g., Einstein field equations, it would be of importance
to investigate whether further extensions of Campbell-Magaard theorem could
be achieved in order to accommodate these models. We are currently doing
some research in this direction.

\section{Acknowledgments}

The authors thank Dr. E. M. Monte for helpful discussions. This work was
supported financially by CNPq.


\begin{references}
\bibitem{string}  Antoniadis, I, Phys. Lett. {\bf B 246}, 377 (1990).
Witten, E, Nucl. Phys {\bf B 471}, 135 (1996). Arkani-Hamed, N, Dimopoulos,
S and Dvali, G. Phys. Lett. {\bf B 436}, 257 (1998).

\bibitem{brane}  Polchinski, J, ``Fields, Strings and Duality'' TASI 1996,
eds. C . Efthimion and B. Greene (World Scientific, Singapure, 1996): 293.

\bibitem{randall}  Randall, L and Sundrum, R, Phys. Rev. Lett. {\bf 83},
3370 (1999). Randall, L and Sundrum, R, Phys. Rev. Lett. {\bf 83}, 4690
(1999).

\bibitem{wesson}  Overduim, J and Wesson, P, Phys. Rep. {\bf 283}, 303
(1997).

\bibitem{einsenhart}  Einsenhart, L, ``Riemannian Geometry'' (Princeton
Univ. Press, 1949), 143.

\bibitem{campbell}  Campbell, J. ``A Course of Differential Geometry''
(Oxford: Claredon, 1926).

\bibitem{magaard}  Magaard, L, ``Zur einbettung riemannscher Raume in
Einstein-Raume und konformeuclidische Raume'' (PhD Thesis, Kiel, 1963)

\bibitem{romero}  Romero, C, Tavakol, R. and Zalaletdinov, R, Gen. Rel.
Grav. {\bf 28}, 365 (1996). Lidsey, J, Romero, C, Tavakol, R, and Rippl, S,
Class. Quantum Grav. {\bf 14}, 865 (1997).

\bibitem{lidsey}  Anderson, E and Lidsey, J, gr-qc/0106090.

\bibitem{fdahia}  Dahia, F, ``Imers\~{a}o do espa\c{c}o-tempo em cinco dimens%
\~{o}es e a generaliza\c{c}\~{a}o do teorema de Campbell-Magaard'' (PhD
Thesis, UFPB, Brazil, 2001).
\end{references}
\end{document}